\documentclass[12pt,preprint]{aastex6}

\usepackage{amsmath}
\usepackage{natbib}
\usepackage{xcolor}
\usepackage[]{graphics}
\usepackage{epsfig}
\usepackage{epstopdf}

\usepackage{mathrsfs}
\begin{document}

\title{Rotating Black Hole Shadow in Perfect Fluid Dark Matter}

\author{
   Xian Hou,\altaffilmark{1,2,3}
   Zhaoyi Xu,\altaffilmark{1,2,3}
  Jiancheng Wang\altaffilmark{1,2,3}
 }

\altaffiltext{1}{Yunnan Observatories, Chinese Academy of Sciences, 396 Yangfangwang, Guandu District, Kunming 650216, P. R. China; {\tt xianhou.astro@gmail.com, zyxu88@ynao.ac.cn, jcwang@ynao.ac.cn}}
\altaffiltext{2}{Key Laboratory for the Structure and Evolution of Celestial Objects, Chinese Academy of Sciences, 396 Yangfangwang, Guandu District, Kunming 650216, P. R. China}
\altaffiltext{3}{Center for Astronomical Mega-Science, Chinese Academy of Sciences, 20A Datun Road, Chaoyang District, Beijing 100012, P. R. China}

%\shorttitle{Rotating Black Hole Shadow in Perfect Fluid Dark Matter}
%\shortauthors{X. Hou et al.}

\begin{abstract}
We study analytically the shadow cast by the rotating black hole in the perfect fluid dark matter. The apparent shape of the shadow depends upon the black hole spin $a$ and the perfect fluid dark matter intensity parameter $k$ ($k>0$). In general, the shadow is a perfect circle in the non-rotating case ($a=0$) and a deformed one in the rotating case ($a\neq 0$). The deformation gets more and more significant with the increasing $a$, similar to the Schwarzschild and Kerr black holes. In addition, there exists a reflection point $k_0$. When $k<k_0$, the size of the shadow decreases with the increasing $k$ and the distortion increases with the increasing $k$. When $k>k_0$, the size of the shadow increases with the increasing $k$ and the distortion decreases with the increasing $k$. Furthermore, the energy emission rate of the black hole in the perfect fluid dark matter increases with the increasing $k$ and the peak of the emission shifts to higher frequencies. Finally, we propose that to observe the effect of the black hole spin $a$ and the perfect fluid dark matter intensity $k$ on the shadow of the black hole  Sgr A$^{*}$ at the center of the Milky Way, highly improved techniques would be necessary for the development of future astronomical instruments.
 
\end{abstract}


\section{INTRODUCTION}
Supermassive black holes are generally believed to reside at the heart of most galaxies including our own galaxy, the Milky Way, though direct detection of a black hole is still one of the most important unresolved problems in astronomy. Among the different methods employed to determine the nature of the black hole, i.e., mass and spin of the black hole, observing the shadow of the black hole remains probably the most exciting and interesting one. The black hole shadow is the optical appearance cast by the black hole when there is a bright distant source behind it. It appears as a two-dimensional dark zone for an distant observer, like us on Earth. The black hole shadow is a natural result of Einstein's theory of General Relativity (GR), so it can not only provide us information on fundamental properties of the black hole, but also serves as a useful tool of testing GR. Attempts to observe the shadow of a black hole are ongoing. For example, the sub-millimeter ``Event Horizon Telescope'' (EHT)\footnote{www.Eventhorizontelescope.org.} \citep{2008Natur.455...78D} based on the very-long baseline interferometry (VLBI) are expected to obtain the first images of the black hole Sgr A$^{*}$ at the center of our own galaxy and of the black hole M87 in the Virgo A galaxy in the near future. 

%and the BlackHoleCam (BHC)\footnote{BlackHoleCam.org.} 

The boundary of the shadow of a non-rotating black hole, like the Schwarzschild black hole, is a perfect circle and was first studies by \cite{1966MNRAS.131..463S} and later by \cite{1979A&A....75..228L} who further considered the effect of a thin accretion disk on the shadow. The shadow of the rotating black hole, i.e., the regular Kerr black hole, is no longer circular but rather deformed \citep{1973blho.conf..215B,2009PhRvD..80b4042H}. The apparent shape of the shadow depends upon the black hole space-time metric. Shadows in various black hole space-time have been examined extensively in the literature in the last decades \citep[e.g.,][]{2000CQGra..17..123D,2005PASJ...57..273T,2018PhRvD..97f4021T,2013JCAP...11..063W,2013Ap&SS.344..429A,2009IJMPD..18..983S,2012PhRvD..85f4019A,2013PhRvD..87d4057A,2012PhLB..711...10B,2013PhRvD..88f4004A,2017JCAP...10..051W,2010CQGra..27t5006B,2016PhRvD..94h4025Y,2017PhLB..768..373C,2016arXiv161009477D,2016PhRvD..93j4004A,2016PhRvD..94b4054A,2018GReGr..50..103S,2014PhRvD..89l4004G,2018PhRvD..97j4062P,2018EPJC...78...91E}. This topic has also been extended to black holes in modified GR \citep[e.g.,][]{2010PhRvD..81l4045A,2017arXiv171209793K,2017CaJPh..95.1299M,2018arXiv180104592V}, to black holes with higher or extra dimensions \citep[e.g.,][]{2014PhRvD..90b4073P, 2015EPJC...75..399A,2017arXiv170709521A,2017arXiv170707125P,2018EPJC...78...91E} and black holes surrounded by plasma \citep[e.g.,][]{2015PhRvD..92h4005A,2015PhRvD..92j4031P}. Multiple shadows of a single black hole or the shadow of multiple black holes have also been discussed recently \citep[e.g.,][]{2015PhRvL.115u1102C,2018PhRvD..97h4024G,2012PhRvD..86j3001Y,2017PhLB..768..373C}. Besides these analytical work on different types of black holes, there have been various analytical and simulation-based work dedicated to the black hole Sgr A$^{*}$ by taking into account more realistic situations such as accretion flow and relativistic jets \citep[e.g.,][]{2000ApJ...528L..13F,2007CQGra..24S.259N,2010ApJ...717.1092D,2014A&A...570A...7M,2015ApJ...799....1C,2016ApJ...820..137B,2017ApJ...837..180G}. Accretion models can thus be constrained by comparing with the EHT observations of Sgr A$^{*}$. The possibility of testing theories of gravity basing on the shadow of Sgr A$^{*}$ has been explored equally in the literature \citep[e.g.,][]{2006JPhCS..54..448B,2009PhRvD..79d3002B,2014ApJ...784....7B,2015ApJ...814..115P,2015MNRAS.454.2423A,2016PhRvL.116c1101J,2018NatAs...2..585M}. More interestingly, \cite{2018arXiv180609415X} suggested that the black hole shadow can be used to determine the matter category, such as dark matter, dust and radiation, around a black hole under the assumption of perfect fluid matter. See \cite{2018GReGr..50...42C} for a recent review of the study of black hole shadows.   

Studying the black hole shadow in the presence of dark matter and dark energy would be of specific interests given that the Universe is domininated by dark mater (27\%) and dark energy (68\%), while the contribution of baryonic matter is minor (5\% to the total mass-energy of the Universe), according to the Standard Model of Cosmology. Recently, the black hole shadow in quintessence has been discussed by \cite{2017arXiv171102898P} and \cite{2017IJMPD..2650051A}, while \cite{2018JCAP...07..015H} studied the shadow of the black hole Sgr A$^{*}$ in the Cold Dark Matter \citep[CDM,][]{1991ApJ...378..496D,1996ApJ...462..563N,1997ApJ...490..493N} and Scalar Field Dark Matter \citep[SFDM. e.g.,][]{2000PhRvL..84.3760S,2002CQGra..19.6259U,2011JCAP...05..022H} halos. The CDM model is the current leading dark matter model despite the fact that it is in tension with long-standing (and more recent) small-scale structure observations \citep{2018PhR...730....1T}. The SFDM model, although less  popular, can provide good concordance with both large-scale and small-scale structure observations. Another alternative dark matter model is the phenomenological Perfect Fluid Dark Matter (PFDM) model \citep[e.g.,][]{2003CQGra..20.1187K,2010PhLB..694...10R}, in which dark matter is described as a perfect fluid. Although simple, this model has the analytical form and the possibility of explaining the asymptotically flat rotation velocity in spiral galaxies \citep[e.g.,][]{2003gr.qc.....3031K,2005CQGra..22..541K,2012PhRvD..86l3015L}. This work will investigate the shadow of rotating black hole in PFDM and discuss its applications to Sgr A$^{*}$, complementary to the work mentioned above.

%Studying the shadow of the black hole Sgr A$^{*}$ in dark matter halo at the center of the Milky Way has particular interest given that the Galactic center is predicted to be the brightest dark matter source in the Universe, due to its proximity and high dark matter density. In addition, evidence of dark matter annihilation signal from the leading dark matter particle candidate, Weakly Interacting Massive Particles (WIMPs), in the form of gamma rays has emerged in the past decade, basing on observations of the Galactic center with space and ground-based high energy telescopes like {\it Fermi}-LAT and H.E.S.S \citep[See][for a review on this topic]{2015arXiv151102031C}. In this context, 

The paper is organized as follows. In Section \ref{metric}, we introduce the space-time metrics for the spherically symmetric and rotating black hole in PFDM. In Section \ref{motion}, we derive the complete null geodesic equations for a test particle moving around the rotating black hole in PFDM. In Section \ref{shadow}, we study the apparent shapes of the shadow cast by the rotating black hole in PFDM. The energy emission rate of the rotating black hole in PFDM is investigated in Section \ref{emission} and we discuss our results in Section \ref{discussion}.

\section{BLACK HOLE SPACE-TIME IN Perfect Fluid Dark Matter}
\label{metric}
\subsection{Case of spherically symmetric black hole}
The spherically symmetric black hole space-time metric in PFDM is \citep{2003gr.qc.....3031K,2012PhRvD..86l3015L}
\begin{equation}
ds^{2}=-f(r)dt^{2}+\frac{dr^{2}}{f(r)}+r^{2}(d\theta^{2}+sin^{2}\theta d\phi^{2}),
\label{SBH1}
\end{equation}
with
\begin{equation}
f(r)=1-\dfrac{2M}{r}+\dfrac{k}{r}\ln\left(\dfrac{r}{\mid k\mid}\right),
\label{fr1}
\end{equation}
where $M$ is the black hole mass and $k$ is a parameter describing the intensity of the PFDM. If the PFDM is absent $(k=0)$, the above space-time metric reduces to the Schwarzschild black hole.

\subsection{Case of rotating black hole}
The rotating black hole space-time metric in PFDM is \citep{2018CQGra..35k5003X}
\begin{equation}
ds^{2}=-\left(1-\dfrac{2Mr-k r ln\left(\dfrac{r}{\mid k\mid}\right)}{\Sigma^{2}}\right)dt^{2}+\dfrac{\Sigma^{2}}{\Delta}dr^{2}-\dfrac{2a sin^{2}\theta\left(2Mr-k r ln\left(\dfrac{r}{\mid k\mid}\right)\right)}{\Sigma^{2}}d\phi dt+\Sigma^{2}d\theta^{2}$$$$
+sin^{2}\theta \left(r^{2}+a^{2}+a^{2}sin^{2}\theta\dfrac{2Mr-k r ln\left(\dfrac{r}{\mid k\mid}\right)}{\Sigma^{2}}\right)d\phi^{2},
\label{KBH1}
\end{equation}
where
\begin{align}
&\Sigma^{2}=r^{2}+a^{2}cos^{2}\theta,\\
&\Delta=r^{2}-2Mr+a^{2}+k r ln\left(\dfrac{r}{\mid k\mid}\right).
\label{KBH2}
\end{align}
If the PFDM is absent $(k=0)$, the above space-time metric reduces to the Kerr black hole.

The values of $k$ can be both positive and negative. Here we consider only positive $k$. The case of negative $k$ can be studied in a similar way as presented in this work.

\section{NULL GEODESICS}
\label{motion}
The shape and size of the shadow depend  upon completely on the geometry of the black hole. It is thus necessary to first study the geodesic structure of a test particle for the above space-time metric (Eq. \ref{KBH1}). To do this, we adopt the Hamilton-Jacobi equation and Carter constant separable method \citep{1968PhRv..174.1559C} to obtain the complete geodesic equations. The general form of the Hamilton-Jacobi equation can be expressed as
\begin{equation}
\frac{\partial S}{\partial \sigma}=-\frac{1}{2}g^{\mu\nu}\frac{\partial S}{\partial {x^{\mu}}}\frac{\partial S}{\partial {x^{\nu}}},
\label{HJ1}
\end{equation}
where $\sigma$ is an affine parameter along the geodesics and $S$ is the Jacobi action of which the separable solution is
\begin{equation}
S=\frac{1}{2}m^2\sigma-Et+L\phi+S_r(r)+S_{\theta}(\theta).
\label{HJ2}
\end{equation}
$m$, $E$ and $L$ are, respectively, the test particle's mass, energy and angular momentum, with respect to the rotation axis. $S_r(r)$ and $S_{\theta}(\theta)$ are, respectively, functions of $r$ and $\theta$. Combining Eq. (\ref{HJ1}) and Eq. (\ref{HJ2}) and applying the variable separable method, we get the null geodesic equations for a test particle around the rotating black hole in PFDM as
\begin{align}
\label{HJ3}
&\Sigma\frac{dt}{d\sigma}=\frac{r^2+a^2}{\Delta}[E(r^2+a^2)-aL]-a(aE\sin^2\theta-L),\\
&\Sigma\frac{dr}{d\sigma}=\sqrt{\mathcal{R}},\\
&\Sigma\frac{d\theta}{d\sigma}=\sqrt{\Theta},\\
\label{HJ4}
&\Sigma\frac{d\phi}{d\sigma}=\frac{a}{\Delta}[E(r^2+a^2)-aL]-\left(aE-\frac{L}{\sin^2\theta}\right),
\end{align}
where $\mathcal{R}(r)$ and $\Theta(\theta)$ read as
\begin{align}
\label{HJ5}
&\mathcal{R}(r)=[E(r^2+a^2)-aL]^2-\Delta[m^2r^2+(aE-L)^2+\mathcal{K}],\\
&\Theta(\theta)=\mathcal{K}-\left(  \dfrac{L^2}{\sin^2\theta}-a^2E^2  \right) \cos^2\theta,
\end{align}
with $\mathcal{K}$ the Carter constant. The dynamics of the test particle around the rotating black hole in PFDM are fully described by the geodesic equations (\ref{HJ3})$-$(\ref{HJ4}). The boundary of the shadow is completely determined by the unstable circular orbit which satisfies the condition
\begin{equation}
\mathcal{R}=\frac{\partial\mathcal{R}}{\partial r}=0.
\label{R1}
\end{equation}
Introducing two impact parameters $\xi$ and $\eta$ as
\begin{equation}
\xi=L/E,  \;\; \;\; \;\;   \eta=\mathcal{K}/E^2,
\end{equation}
and considering the case of photons ($m=0$), for an observer at the infinity (photons will arrive near the equatorial plane with $\theta=\pi/2$), we obtain
\begin{align}
\label{R2}
&(r^2+a^2-a\xi)^2-[\eta+(\xi-a)^2](r^2f(r)+a^2)=0,\\
&4r(r^2+a^2-a\xi)-[\eta+(\xi-a)^2](2rf(r)+r^2f'(r))=0.
\label{R3}
\end{align}
From Eqs. (\ref{R2}) and (\ref{R3}), we can get the expressions of $\xi$ and $\eta$ as
\begin{align}
&\xi=\frac{(r^2+a^2)(rf'(r)+2f(r))-4(r^2f(r)+a^2)}{a(rf'(r)+2f(r))},\\
&\eta=\frac{r^3[8a^2f'(r)-r(rf'(r)-2f(r))^2]}{a^2(rf'(r)+2f(r))^2}.
\end{align}
Furthermore, we have
\begin{align}
\xi^2+\eta &=2r^2+a^2+\frac{16(r^2f(r)+a^2)}{(rf'(r)+2f(r))^2}-\frac{8(r^2f(r)+a^2)}{rf'(r)+2f(r)}\\
                &=2r^2+a^2+\frac{8\Delta[2-(rf'(r)+2f(r)]}{(rf'(r)+2f(r))^2},
\end{align}
where 
\begin{equation}
f'(r)=\dfrac{2M+k}{r^2}-\dfrac{k}{r^2}\ln\left(\dfrac{r}{\mid k \mid}\right).
\label{dFR}
\end{equation}

\section{BLACK HOLE SHADOW}
\label{shadow}
In order to study the shape of the black hole shadow, we introduce the celestial coordinates $\alpha$ and $\beta$ as 
\begin{align}
& \alpha = \lim_{r_o\to \infty}\left( -r_o^2 \sin \theta_o \dfrac{d\phi}{dr}  \right),\\
& \beta = \lim_{r_o \to \infty}\left( r_o^2 \dfrac{d\theta}{dr}  \right).
\end{align}
Here we assume the observer is at infinity. $r_o$ is the distance between the observer and the black hole, and $\theta_o$ is the inclination angle. $\alpha$ and $\beta$ are the apparent perpendicular distances of the shadow as seen from, respectively, the axis of symmetry, and its projection on the equatorial plane.

Using the null geodesic equations (\ref{HJ3})$-$(\ref{HJ4}), we get the relations between celestial coordinates and impact parameters as
\begin{align}
& \alpha = -\dfrac{\xi}{sin \theta},\\
& \beta = \pm \sqrt{\eta + a^2\cos^2\theta -\xi^2\cot^2 \theta}.
\end{align}
In the equatorial plane ($\theta=\pi/2$), $\alpha$ and $\beta$ reduce to
\begin{align}
& \alpha = -\xi,\\
& \beta = \pm \sqrt{\eta }.
\end{align}

\begin{figure}[htbp!]
  \centering
   \includegraphics[scale=0.35]{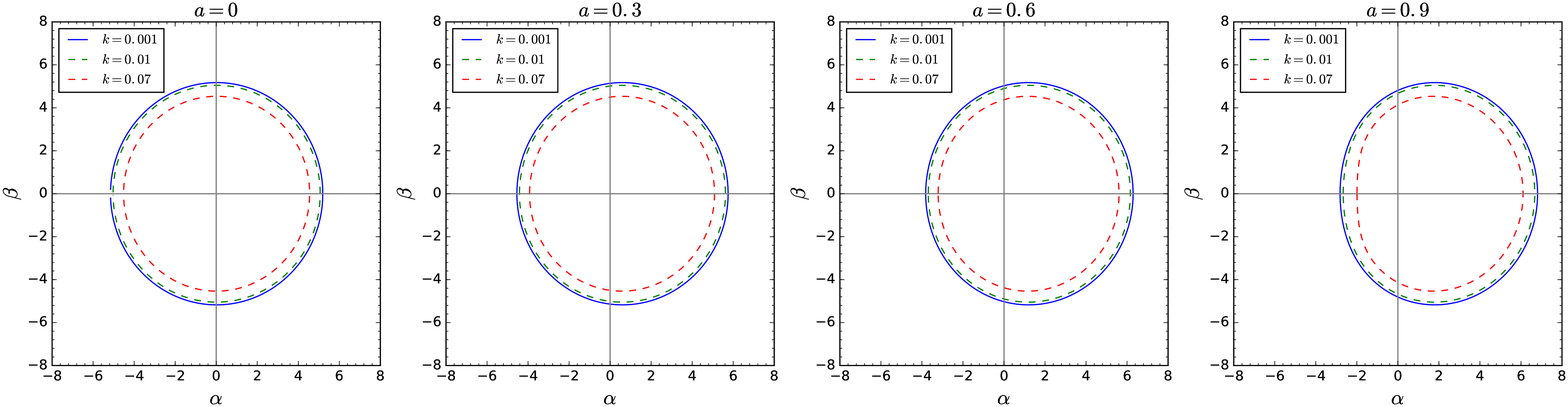}
   \includegraphics[scale=0.35]{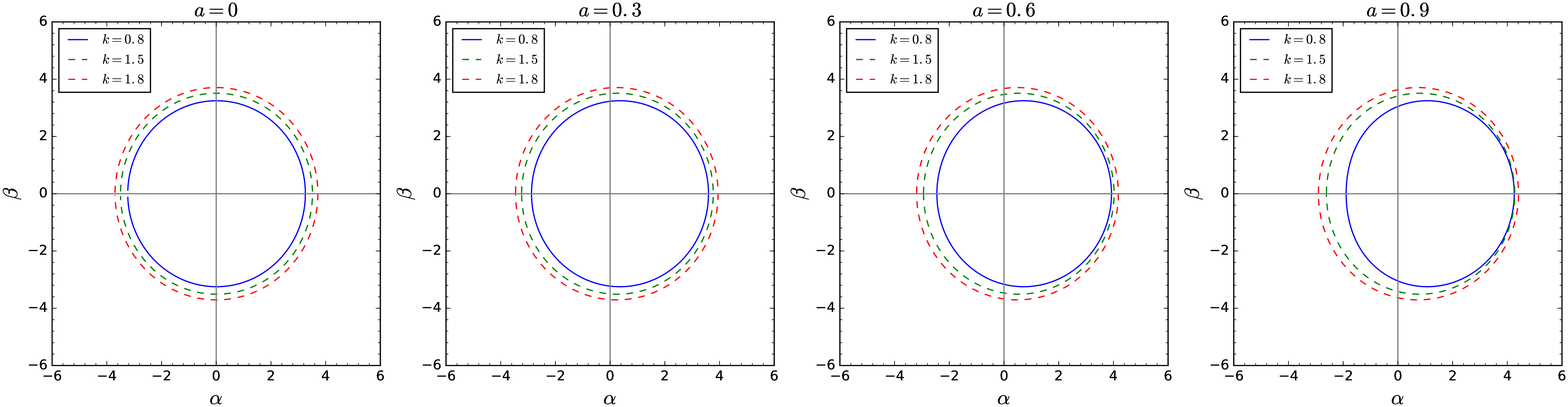}
   \caption{Silhouette of the shadow cast by the rotating black hole in perfect fluid dark matter for different values of the parameters $a$ and $k$ ($k>0$). Upper panel: $k<k_0$; lower panel: $k>k_0$.}
  \label{shadow_1}
\end{figure}

Different shapes of the shadow can be obtained by plotting $\beta$ against $\alpha$. Through calculations, we find $f(r)$ has a reflection point at $k_0=2/(1+e)$. When $k<k_0$, $f(r)$ monotonically decreases with $k$, while when $k>k_0$, $f(r)$ monotonically increases with $k$. We thus plot different shadows in Figure \ref{shadow_1} by choosing reasonable values of $k$ on both sides of the reflection point. When $k<k_0$, the shadow is a perfect circle and the size decreases with the increasing $k$ in the non-rotating case ($a=0$),  while in the rotating case ($a\neq 0$), the shadow gets more and more distorted with the increasing $a$, and the size decreases with the increasing $k$, similar to the case of $a=0$. When $k>k_0$, the structure of the shadow is similar to what we find for $k<k_0$ except that the size of the shadow instead increases with the increasing $k$.

%\begin{figure}[htbp]
%  \centering
%   \includegraphics[scale=0.35]{shadow_3}
%   \includegraphics[scale=0.35]{shadow_4}
%   \caption{Silhouette of the shadow cast by the rotating black hole in perfect fluid dark matter for different values of parameters $a$ and $k$ ($k<0$).}
%  \label{shadow_2}
%\end{figure}

To extract detailed information from the shadow and connect to astronomical observations, we adopt the two observables defined in \cite{2009PhRvD..80b4042H}: the radius of the shadow $R_s$ which approximately describes the size of the shadow and $\delta_s$ which measures its deformation \citep[Figure 3,][]{2018JCAP...07..015H}. $R_s$ is defined as the radius of a reference circle passing through three points: $A(\alpha_r, 0)$, $B(\alpha_t, \beta_t)$ and $D(\alpha_b, \beta_b)$. From the geometry of the shadow, $R_s$ can be calculated as
%The points $C(\alpha_p, 0)$ and $F(\tilde{\alpha}_p, 0)$ are where the circle of the shadow and the reference circle cut the horizontal axis at the opposite side of $A(\alpha_r, 0)$, respectively. $d_s$ is the distance from the most left position (C) of the shadow to the reference circle (F). $R_s$ gives  with respect to the reference circle.
\begin{equation}
R_s = \dfrac{(\alpha_t-\alpha_r)^2 + \beta_t^2}{2|\alpha_r-\alpha_t|},
\end{equation}
and $\delta_s$ can be expressed as
\begin{equation}
\delta_s = \dfrac{d_s}{R_s} = \dfrac{|\alpha_p-\tilde{\alpha}_p|}{R_s},
\end{equation}
where $d_s$ is the distance between the most left position of the shadow and of the reference circle.

%Considering the relation $\tilde{\alpha}_p=\alpha_r-2 R_s$, we have
%\begin{equation}
%\delta_s = 2-\dfrac{D_s}{R_s}
%\end{equation}
%where $D_s=\alpha_r-\alpha_p$ is the diameter of the shadow along the axis of $\alpha$.

%\begin{figure}[htbp]
%  \centering
%  \includegraphics[scale=0.5]{shadow_ref}
%   \caption{Schematic illustration of the black hole shadow and the reference circle.}
%      %\caption{Schematic illustration of the black hole shadow and the reference circle \citep{2016PhRvD..93j4004A}.}
%  \label{shadow_ref}
%\end{figure}

For non-rotating black holes ($a=0$), the shadow is a perfect circle with radius of $R_s$. So
\begin{equation}
\alpha^2 + \beta^2 = \xi^2 +\eta= R_s^2. 
\end{equation}

\begin{figure}[htbp]
  \centering
   \includegraphics[scale=0.5]{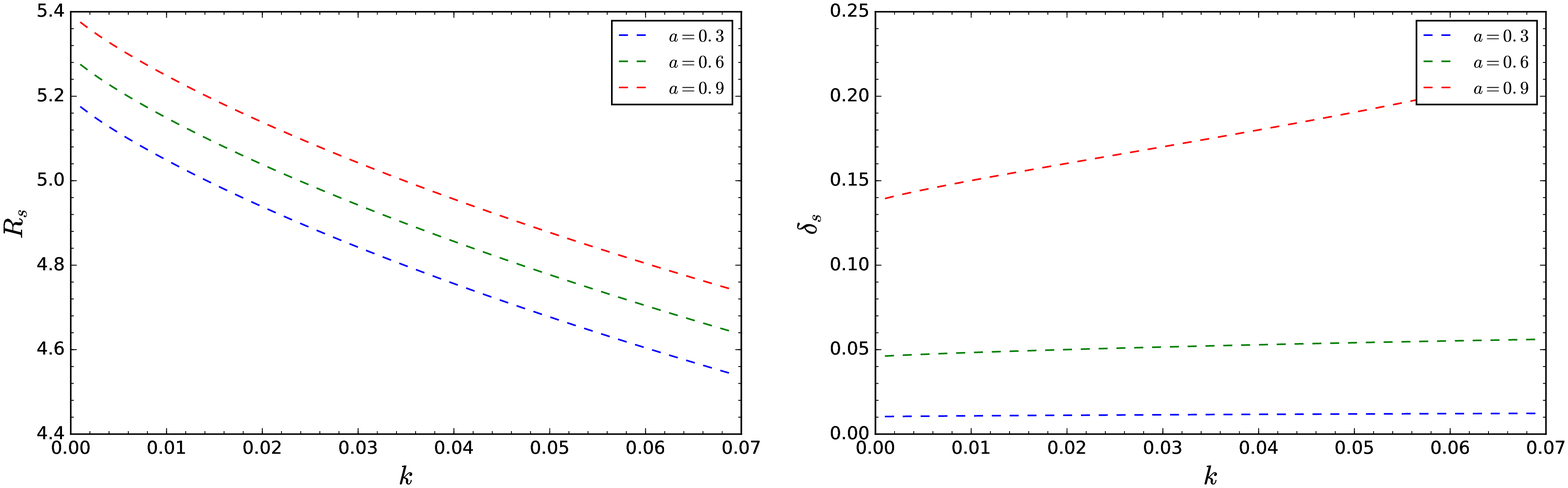}
   \includegraphics[scale=0.5]{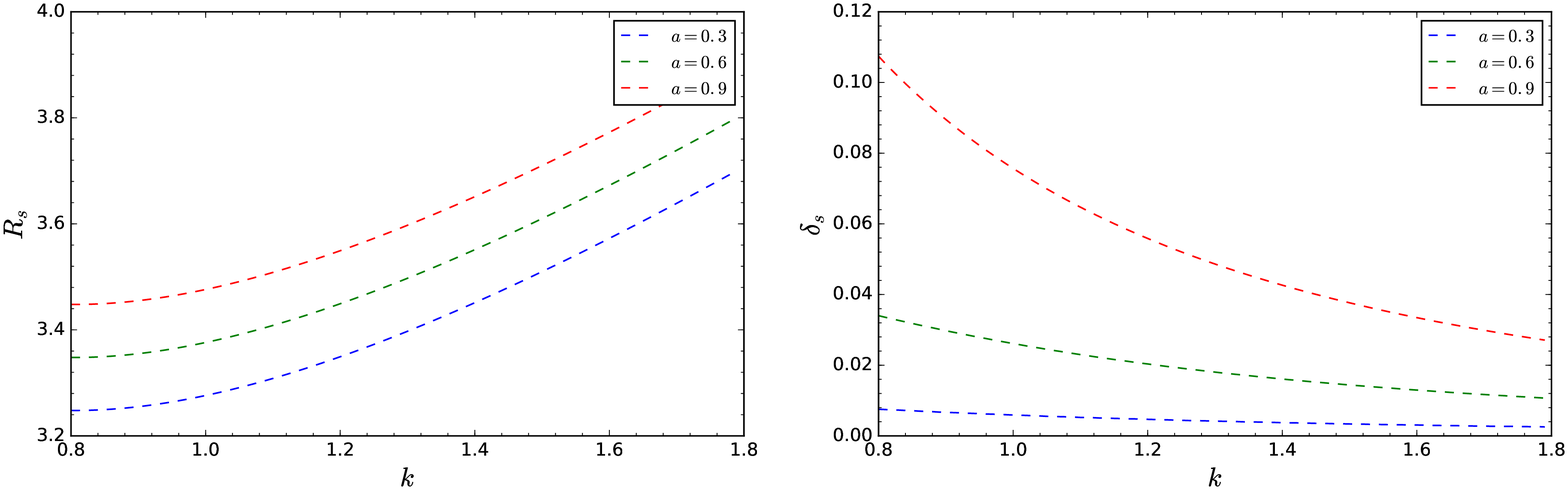}
   \caption{Variation of the radius $R_s$ (left) and the distortion parameter $\delta_s$ (right) of the black hole shadow with the parameters $a$ and $k$ ($k>0$). Upper panel: $k<k_0$; lower panel: $k>k_0$. The lines of $R_s$ have been moved up vertically to visualize the trend of $R_s$ for different $a$ by adding a constant to $R_s$.}
  \label{Rs}
\end{figure}

%\begin{figure}[htbp]
%  \centering
%   \includegraphics[scale=0.5]{Rs_deltaS_3}
%   \includegraphics[scale=0.5]{Rs_deltaS_4}
%   \caption{Variation of the radius $R_s$ (left) and the distortion parameter $\delta_s$ (right) of the black hole shadow with the parameters $a$ and $k$ ($k<0$). The lines of $R_s$ have been moved up vertically to visualize the trend of $R_s$ for different $a$ by adding a constant to $R_s$.}
%  \label{Rs}
%\end{figure}

We show in Figure \ref{Rs} the evolution of the radius $R_s$ and the distortion parameter $\delta_s$ with the parameters $a$ and $k$ on both sides of the reflection point $k_0$. We find that when $k<k_0$, $R_s$ decreases with the increasing $k$, but almost does not vary with $a$ (a constant has been added to visualize the trend of $R_s$ for different $a$), while $\delta_s$ increases monotonically with the increasing $k$ and $a$.  When $k>k_0$, the evolution pattern is contrary to that of $k<k_0$. This result is consistent with what can be inferred from Figure \ref{shadow_1}.

\section{ENERGY EMISSION RATE}
\label{emission}
We assume that, for an observer located at infinity, the black hole shadow approaches to the high energy absorption cross section of the black hole. For a spherically symmetric black hole, the high energy absorption cross section oscillates around a limiting constant value $\sigma_{lim}$ which is approximately equal to the geometrical cross section of the photon sphere \citep{1973PhRvD...7.2807M,1973grav.book.....M} and can be expressed as \citep{2013JCAP...11..063W} 
\begin{equation}
\sigma_{lim} \approx \pi R_s^2,
\end{equation}
where $R_s$ is the radius of the black hole shadow. We extend this result to the rotating black hole considered in this work, given that the shadow resembles a standard circle even for extreme values of $a$ and $k$ (Figure \ref{shadow_1}). The high energy emission rate of the black hole reads as 
\begin{equation}
\dfrac{d^2E(\omega)}{d\omega dt} = \dfrac{2\pi^2\sigma_{lim} }{e^{\omega/T}-1}\omega^3,
\end{equation}
where $\omega$ is the frequency of photon and $T$ is the Hawking temperature for the outer event horizon ($r_+$) which can be expressed as
%\begin{equation}
%T = \lim_{\theta=0,  r \to r_+} \dfrac{\partial_r \sqrt{g_{tt}}   }{2\pi \sqrt{g_{rr}}}
%\end{equation}
%From Eq. (\ref{KBH1}), we have for the rotating black hole in dark matter halo
%\begin{equation}
%g_{tt} = 1-\dfrac{r^{2}-f(r)r^{2}}{\Sigma^{2}},  ~~~~~ g_{rr} = \dfrac{\Sigma^{2}}{\Delta}
%\end{equation}
\begin{equation}
T = \dfrac{r_+^2f'(r_+)(r_+^2+a^2) + 2a^2r_+(f(r_+)-1)  }{ 4\pi (r_+^2+a^2)^2      }.
\label{T_hawk}
\end{equation} 

If the PFDM is absent ($k=0$), Eq. (\ref{T_hawk}) reduces to the regular Kerr black hole 
\begin{equation}
T_{\rm Kerr} = \dfrac{r_+^2-a^2 }{ 4\pi r_+ (r_+^2+a^2) },
\end{equation} 
where $r_+ = M+\sqrt{M^2-a^2}$.
%defined as the greater root of the solution for $1/g_{rr}=0$.

\begin{figure}[htbp]
  \centering
   \includegraphics[scale=0.5]{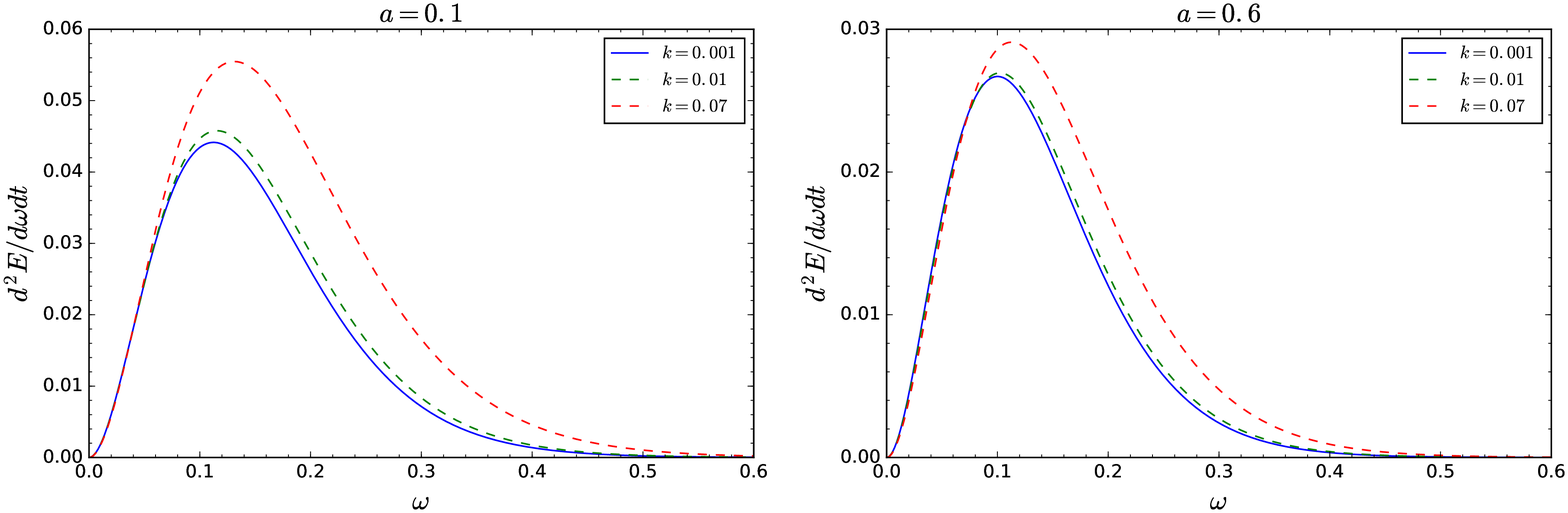}
   \includegraphics[scale=0.5]{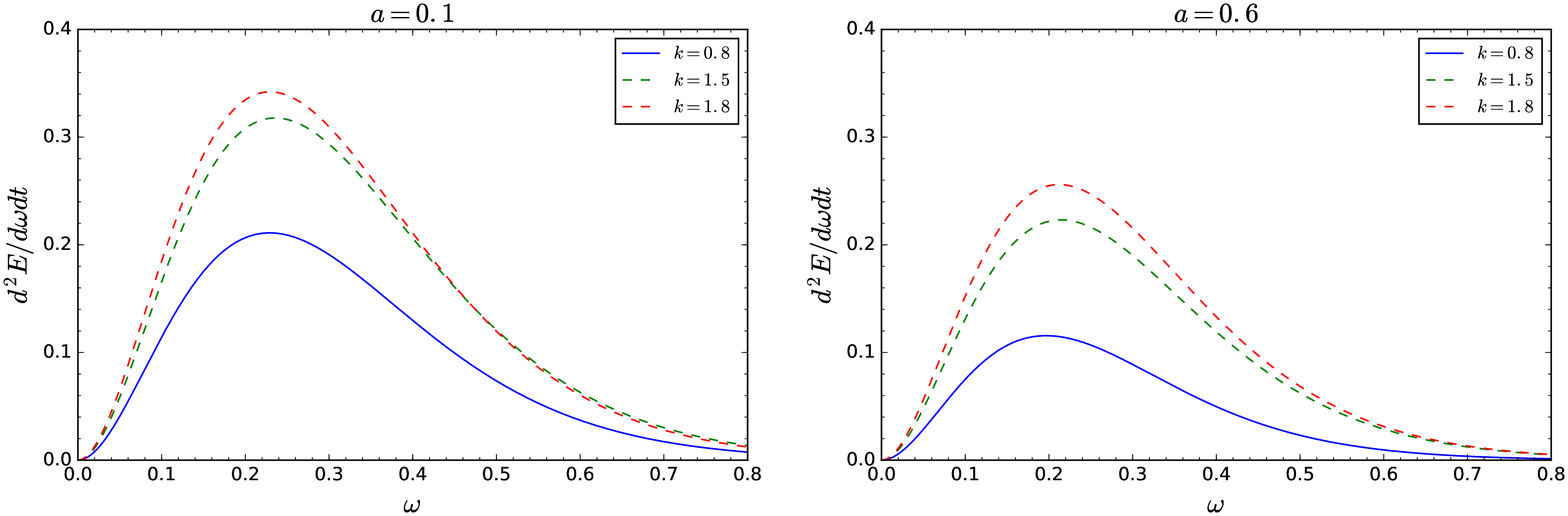}
   \caption{Evolution of the black hole energy emission rate with the frequency $\omega$ for different values of the parameters $a$ and $k$ ($k>0$). Upper panel: $k<k_0$; lower panel: $k>k_0$.}
  \label{Erate_1}
\end{figure}

The energy emission rate evolution with the photon frequency $\omega$ is shown in Figure \ref{Erate_1} for different values of the parameters $a$ and $k$ on both sides of the reflection point $k_0$.  It is clear that the peak of the emission increases with the increasing $k$ and shifts to higher frequencies regardless of the range of $k$ specified.

%\begin{figure}[htbp]
%  \centering
%   \includegraphics[scale=0.5]{emiss_rate_3}
%   \includegraphics[scale=0.5]{emiss_rate_4}
%   \caption{Evolution of the emission rate with the frequency $\omega$ of the black hole for different values of the parameters $a$ and $k$ ($k<0$). }
%  \label{Erate_2}
%\end{figure}

\section{DISCUSSION}
\label{discussion}
In this work, we study the rotating black hole shadow in the perfect fluid dark matter by investigating how the shadow varies with the black hole spin $a$ and the perfect fluid dark matter intensity $k$ ($k>0$). We show that for an observer located at an infinite distance and in the equatorial plane of the black hole, the shadow is a perfect circle in the non-rotating case ($a=0$) and a deformed one in the rotating case ($a\neq 0$). The evolution of the shadow size $R_s$ and the distortion parameter $\delta_s$ is dependant of the range of $k$ specified. When $k$ is smaller than the reflection point $k_0$, $R_s$ decreases with the increasing $k$ and $\delta_s$ increases with the increasing $k$, while when $k>k_0$, instead, $R_s$ increases with the increasing $k$ and $\delta_s$ decreases with the increasing $k$. On the other hand, the shadow gets more and more distorted with the increasing $a$ while the size does not vary significantly. Under the assumption that the black hole shadow equals to the high energy absorption cross section, we further calculate the energy emission rate of the black hole in the perfect fluid dark matter. Independent of the range of $k$ chosen, the emission rate increases with the increasing $k$ and the peak shifts to higher frequencies. 

In addition, we can estimate the angular radius of the shadow as $\theta_s = R_s M/D$, where $M$ is the black hole mass and $D$ is the distance between the black hole and the observer. \cite{2012PhRvD..85f4019A} further proposed to estimate the angular radius as $\theta_s = 9.87098\times 10^{-6} R_s(M/\rm M\odot)(1 \rm{kpc}/\it{D})$ $\,\mu$as with $\rm M\odot$ the solar mass. Taking the supermassive black hole Sgr A$^{*}$ at the center of the Milky Way as example, $M=4.3\times10^6\, \rm M\odot$ and $D=8.3$ kpc. We find that to extract the information of the perfect fluid dark matter intensity parameter $k$, an angular resolution of 1 $\mu$as will be enough, while for the black hole spin $a$, a resolution of much less than 1 $\mu$as will be needed. Nevertheless, both resolutions required are out of the capacity of the current astronomical instruments. Currently, EHT has a resolution of $\sim$ 60 $\mu$as at 230 GHz and is expected to achieve 15 $\mu$as by observing at a higher frequency of 345 GHz and adding more VLBI telescopes. The space-based VLBI RadioAstron \citep{2013ARep...57..153K}\footnote{http://www.asc.rssi.ru/radioastron/index.html} will obtain a resolution of $\sim 1-10$ $\mu$as in the future. We conclude that future observations with highly improved techniques would be able to achieve the resolution required to observe the perfect fluid dark matter influence on the shadow of the black hole Sgr A$^{*}$. 

When submitting the paper, we were aware of a similar work by \cite{2018arXiv181004103H}, in which the complete black hole solution in PFDM with a cosmological constant reported by \cite{2018CQGra..35k5003X} were explored. We notice that when the cosmological constant is not considered, their result on the black hole shadow is consistent with ours. They in addition studied the effects of the PFDM parameter and the cosmological constant on the deflection angle of light,  while we investigated in detail the evolution of the shadow as well as the black hole emission rate.

\acknowledgments
We acknowledge the anonymous referee for a constructive report that has significantly improved this paper. We acknowledge the financial support from the National Natural Science Foundation of China under grants No. 11503078, 11573060 and 11661161010.

\bibliography{shadow_BH_perfect_fluid_DM}

\begin{thebibliography}{}
\expandafter\ifx\csname natexlab\endcsname\relax\def\natexlab#1{#1}\fi

\bibitem[{{Abdujabbarov} {et~al.}(2016){Abdujabbarov}, {Amir}, {Ahmedov}, \&
  {Ghosh}}]{2016PhRvD..93j4004A}
{Abdujabbarov}, A., {Amir}, M., {Ahmedov}, B., \& {Ghosh}, S.~G. 2016, \prd,
  93, 104004

\bibitem[{{Abdujabbarov} {et~al.}(2015{\natexlab{a}}){Abdujabbarov},
  {Atamurotov}, {Dadhich}, {Ahmedov}, \&
  {Stuchl{\'{\i}}k}}]{2015EPJC...75..399A}
{Abdujabbarov}, A., {Atamurotov}, F., {Dadhich}, N., {Ahmedov}, B., \&
  {Stuchl{\'{\i}}k}, Z. 2015{\natexlab{a}}, European Physical Journal C, 75,
  399

\bibitem[{{Abdujabbarov} {et~al.}(2013){Abdujabbarov}, {Atamurotov},
  {Kucukakca}, {Ahmedov}, \& {Camci}}]{2013Ap&SS.344..429A}
{Abdujabbarov}, A., {Atamurotov}, F., {Kucukakca}, Y., {Ahmedov}, B., \&
  {Camci}, U. 2013, \apss, 344, 429

\bibitem[{{Abdujabbarov} {et~al.}(2017){Abdujabbarov}, {Toshmatov},
  {Stuchl{\'{\i}}k}, \& {Ahmedov}}]{2017IJMPD..2650051A}
{Abdujabbarov}, A., {Toshmatov}, B., {Stuchl{\'{\i}}k}, Z., \& {Ahmedov}, B.
  2017, International Journal of Modern Physics D, 26, 1750051

\bibitem[{{Abdujabbarov} {et~al.}(2015{\natexlab{b}}){Abdujabbarov},
  {Rezzolla}, \& {Ahmedov}}]{2015MNRAS.454.2423A}
{Abdujabbarov}, A.~A., {Rezzolla}, L., \& {Ahmedov}, B.~J. 2015{\natexlab{b}},
  \mnras, 454, 2423

\bibitem[{{Amarilla} \& {Eiroa}(2012)}]{2012PhRvD..85f4019A}
{Amarilla}, L., \& {Eiroa}, E.~F. 2012, \prd, 85, 064019

\bibitem[{{Amarilla} \& {Eiroa}(2013)}]{2013PhRvD..87d4057A}
---. 2013, \prd, 87, 044057

\bibitem[{{Amarilla} {et~al.}(2010){Amarilla}, {Eiroa}, \&
  {Giribet}}]{2010PhRvD..81l4045A}
{Amarilla}, L., {Eiroa}, E.~F., \& {Giribet}, G. 2010, \prd, 81, 124045

\bibitem[{{Amir} \& {Ghosh}(2016)}]{2016PhRvD..94b4054A}
{Amir}, M., \& {Ghosh}, S.~G. 2016, \prd, 94, 024054

\bibitem[{{Amir} {et~al.}(2017){Amir}, {Pratap Singh}, \&
  {Ghosh}}]{2017arXiv170709521A}
{Amir}, M., {Pratap Singh}, B., \& {Ghosh}, S.~G. 2017, ArXiv e-prints,
  arXiv:1707.09521

\bibitem[{{Atamurotov} {et~al.}(2013){Atamurotov}, {Abdujabbarov}, \&
  {Ahmedov}}]{2013PhRvD..88f4004A}
{Atamurotov}, F., {Abdujabbarov}, A., \& {Ahmedov}, B. 2013, \prd, 88, 064004

\bibitem[{{Atamurotov} {et~al.}(2015){Atamurotov}, {Ahmedov}, \&
  {Abdujabbarov}}]{2015PhRvD..92h4005A}
{Atamurotov}, F., {Ahmedov}, B., \& {Abdujabbarov}, A. 2015, \prd, 92, 084005

\bibitem[{{Bambi} {et~al.}(2012){Bambi}, {Caravelli}, \&
  {Modesto}}]{2012PhLB..711...10B}
{Bambi}, C., {Caravelli}, F., \& {Modesto}, L. 2012, Physics Letters B, 711, 10

\bibitem[{{Bambi} \& {Freese}(2009)}]{2009PhRvD..79d3002B}
{Bambi}, C., \& {Freese}, K. 2009, \prd, 79, 043002

\bibitem[{{Bambi} \& {Yoshida}(2010)}]{2010CQGra..27t5006B}
{Bambi}, C., \& {Yoshida}, N. 2010, Classical and Quantum Gravity, 27, 205006

\bibitem[{{Bardeen}(1973)}]{1973blho.conf..215B}
{Bardeen}, J.~M. 1973, in Black Holes (Les Astres Occlus), ed. C.~{Dewitt} \&
  B.~S. {Dewitt}, 215--239

\bibitem[{{Broderick} {et~al.}(2014){Broderick}, {Johannsen}, {Loeb}, \&
  {Psaltis}}]{2014ApJ...784....7B}
{Broderick}, A.~E., {Johannsen}, T., {Loeb}, A., \& {Psaltis}, D. 2014, \apj,
  784, 7

\bibitem[{{Broderick} \& {Loeb}(2006)}]{2006JPhCS..54..448B}
{Broderick}, A.~E., \& {Loeb}, A. 2006, in Journal of Physics Conference
  Series, Vol.~54, Journal of Physics Conference Series, ed. R.~{Sch{\"o}del},
  G.~C. {Bower}, M.~P. {Muno}, S.~{Nayakshin}, \& T.~{Ott}, 448--455

\bibitem[{{Broderick} {et~al.}(2016){Broderick}, {Fish}, {Johnson},
  {Rosenfeld}, {Wang}, {Doeleman}, {Akiyama}, {Johannsen}, \&
  {Roy}}]{2016ApJ...820..137B}
{Broderick}, A.~E., {Fish}, V.~L., {Johnson}, M.~D., {et~al.} 2016, \apj, 820,
  137

\bibitem[{{Carter}(1968)}]{1968PhRv..174.1559C}
{Carter}, B. 1968, Physical Review, 174, 1559

\bibitem[{{Chan} {et~al.}(2015){Chan}, {Psaltis}, {{\"O}zel}, {Narayan}, \&
  {Sa{\c d}owski}}]{2015ApJ...799....1C}
{Chan}, C.-K., {Psaltis}, D., {{\"O}zel}, F., {Narayan}, R., \& {Sa{\c
  d}owski}, A. 2015, \apj, 799, 1

\bibitem[{{Cunha} \& {Herdeiro}(2018)}]{2018GReGr..50...42C}
{Cunha}, P.~V.~P., \& {Herdeiro}, C.~A.~R. 2018, General Relativity and
  Gravitation, 50, 42

\bibitem[{{Cunha} {et~al.}(2017){Cunha}, {Herdeiro}, {Kleihaus}, {Kunz}, \&
  {Radu}}]{2017PhLB..768..373C}
{Cunha}, P.~V.~P., {Herdeiro}, C.~A.~R., {Kleihaus}, B., {Kunz}, J., \& {Radu},
  E. 2017, Physics Letters B, 768, 373

\bibitem[{{Cunha} {et~al.}(2015){Cunha}, {Herdeiro}, {Radu}, \&
  {R{\'u}narsson}}]{2015PhRvL.115u1102C}
{Cunha}, P.~V.~P., {Herdeiro}, C.~A.~R., {Radu}, E., \& {R{\'u}narsson}, H.~F.
  2015, Physical Review Letters, 115, 211102

\bibitem[{{Dastan} {et~al.}(2016){Dastan}, {Saffari}, \&
  {Soroushfar}}]{2016arXiv161009477D}
{Dastan}, S., {Saffari}, R., \& {Soroushfar}, S. 2016, ArXiv e-prints,
  arXiv:1610.09477

\bibitem[{{de Vries}(2000)}]{2000CQGra..17..123D}
{de Vries}, A. 2000, Classical and Quantum Gravity, 17, 123

\bibitem[{{Dexter} {et~al.}(2010){Dexter}, {Agol}, {Fragile}, \&
  {McKinney}}]{2010ApJ...717.1092D}
{Dexter}, J., {Agol}, E., {Fragile}, P.~C., \& {McKinney}, J.~C. 2010, \apj,
  717, 1092

\bibitem[{{Doeleman} {et~al.}(2008){Doeleman}, {Weintroub}, {Rogers},
  {Plambeck}, {Freund}, {Tilanus}, {Friberg}, {Ziurys}, {Moran}, {Corey},
  {Young}, {Smythe}, {Titus}, {Marrone}, {Cappallo}, {Bock}, {Bower},
  {Chamberlin}, {Davis}, {Krichbaum}, {Lamb}, {Maness}, {Niell}, {Roy},
  {Strittmatter}, {Werthimer}, {Whitney}, \& {Woody}}]{2008Natur.455...78D}
{Doeleman}, S.~S., {Weintroub}, J., {Rogers}, A.~E.~E., {et~al.} 2008, \nat,
  455, 78

\bibitem[{{Dubinski} \& {Carlberg}(1991)}]{1991ApJ...378..496D}
{Dubinski}, J., \& {Carlberg}, R.~G. 1991, \apj, 378, 496

\bibitem[{{Eiroa} \& {Sendra}(2018)}]{2018EPJC...78...91E}
{Eiroa}, E.~F., \& {Sendra}, C.~M. 2018, European Physical Journal C, 78, 91

\bibitem[{{Falcke} {et~al.}(2000){Falcke}, {Melia}, \&
  {Agol}}]{2000ApJ...528L..13F}
{Falcke}, H., {Melia}, F., \& {Agol}, E. 2000, \apjl, 528, L13

\bibitem[{{Gold} {et~al.}(2017){Gold}, {McKinney}, {Johnson}, \&
  {Doeleman}}]{2017ApJ...837..180G}
{Gold}, R., {McKinney}, J.~C., {Johnson}, M.~D., \& {Doeleman}, S.~S. 2017,
  \apj, 837, 180

\bibitem[{{Grenzebach} {et~al.}(2014){Grenzebach}, {Perlick}, \&
  {L{\"a}mmerzahl}}]{2014PhRvD..89l4004G}
{Grenzebach}, A., {Perlick}, V., \& {L{\"a}mmerzahl}, C. 2014, \prd, 89, 124004

\bibitem[{{Grover} {et~al.}(2018){Grover}, {Kunz}, {Nedkova}, {Wittig}, \&
  {Yazadjiev}}]{2018PhRvD..97h4024G}
{Grover}, J., {Kunz}, J., {Nedkova}, P., {Wittig}, A., \& {Yazadjiev}, S. 2018,
  \prd, 97, 084024

\bibitem[{{Harko}(2011)}]{2011JCAP...05..022H}
{Harko}, T. 2011, \jcap, 5, 022

\bibitem[{{Haroon} {et~al.}(2018){Haroon}, {Jamil}, {Jusufi}, {Lin}, \&
  {Mann}}]{2018arXiv181004103H}
{Haroon}, S., {Jamil}, M., {Jusufi}, K., {Lin}, K., \& {Mann}, R.~B. 2018,
  ArXiv e-prints, arXiv:1810.04103

\bibitem[{{Hioki} \& {Maeda}(2009)}]{2009PhRvD..80b4042H}
{Hioki}, K., \& {Maeda}, K.-I. 2009, \prd, 80, 024042

\bibitem[{{Hou} {et~al.}(2018){Hou}, {Xu}, {Zhou}, \&
  {Wang}}]{2018JCAP...07..015H}
{Hou}, X., {Xu}, Z., {Zhou}, M., \& {Wang}, J. 2018, \jcap, 7, 015

\bibitem[{{Johannsen} {et~al.}(2016){Johannsen}, {Broderick}, {Plewa},
  {Chatzopoulos}, {Doeleman}, {Eisenhauer}, {Fish}, {Genzel}, {Gerhard}, \&
  {Johnson}}]{2016PhRvL.116c1101J}
{Johannsen}, T., {Broderick}, A.~E., {Plewa}, P.~M., {et~al.} 2016, Physical
  Review Letters, 116, 031101

\bibitem[{{Kardashev} {et~al.}(2013){Kardashev}, {Khartov}, {Abramov},
  {Avdeev}, {Alakoz}, {Aleksandrov}, {Ananthakrishnan}, {Andreyanov},
  {Andrianov}, {Antonov}, {Artyukhov}, {Arkhipov}, {Baan}, {Babakin},
  {Babyshkin}, {Bartel'}, {Belousov}, {Belyaev}, {Berulis}, {Burke},
  {Biryukov}, \& {Bubnov}}]{2013ARep...57..153K}
{Kardashev}, N.~S., {Khartov}, V.~V., {Abramov}, V.~V., {et~al.} 2013,
  Astronomy Reports, 57, 153

\bibitem[{{Kiselev}(2003{\natexlab{a}})}]{2003CQGra..20.1187K}
{Kiselev}, V.~V. 2003{\natexlab{a}}, Classical and Quantum Gravity, 20, 1187

\bibitem[{{Kiselev}(2003{\natexlab{b}})}]{2003gr.qc.....3031K}
---. 2003{\natexlab{b}}, ArXiv General Relativity and Quantum Cosmology
  e-prints, gr-qc/0303031

\bibitem[{{Kiselev}(2005)}]{2005CQGra..22..541K}
---. 2005, Classical and Quantum Gravity, 22, 541

\bibitem[{{Kumar} {et~al.}(2017){Kumar}, {Pratap Singh}, {Sabir Ali}, \&
  {Ghosh}}]{2017arXiv171209793K}
{Kumar}, R., {Pratap Singh}, B., {Sabir Ali}, M., \& {Ghosh}, S.~G. 2017, ArXiv
  e-prints, arXiv:1712.09793

\bibitem[{{Li} \& {Yang}(2012)}]{2012PhRvD..86l3015L}
{Li}, M.-H., \& {Yang}, K.-C. 2012, \prd, 86, 123015

\bibitem[{{Luminet}(1979)}]{1979A&A....75..228L}
{Luminet}, J.-P. 1979, \aap, 75, 228

\bibitem[{{Mashhoon}(1973)}]{1973PhRvD...7.2807M}
{Mashhoon}, B. 1973, \prd, 7, 2807

\bibitem[{{Misner} {et~al.}(1973){Misner}, {Thorne}, \&
  {Wheeler}}]{1973grav.book.....M}
{Misner}, C.~W., {Thorne}, K.~S., \& {Wheeler}, J.~A. 1973, {Gravitation}

\bibitem[{{Mizuno} {et~al.}(2018){Mizuno}, {Younsi}, {Fromm}, {Porth}, {De
  Laurentis}, {Olivares}, {Falcke}, {Kramer}, \&
  {Rezzolla}}]{2018NatAs...2..585M}
{Mizuno}, Y., {Younsi}, Z., {Fromm}, C.~M., {et~al.} 2018, Nature Astronomy, 2,
  585

\bibitem[{{Mo{\'s}cibrodzka} {et~al.}(2014){Mo{\'s}cibrodzka}, {Falcke},
  {Shiokawa}, \& {Gammie}}]{2014A&A...570A...7M}
{Mo{\'s}cibrodzka}, M., {Falcke}, H., {Shiokawa}, H., \& {Gammie}, C.~F. 2014,
  \aap, 570, A7

\bibitem[{{Mureika} \& {Varieschi}(2017)}]{2017CaJPh..95.1299M}
{Mureika}, J.~R., \& {Varieschi}, G.~U. 2017, Canadian Journal of Physics, 95,
  1299

\bibitem[{{Navarro} {et~al.}(1996){Navarro}, {Frenk}, \&
  {White}}]{1996ApJ...462..563N}
{Navarro}, J.~F., {Frenk}, C.~S., \& {White}, S.~D.~M. 1996, \apj, 462, 563

\bibitem[{{Navarro} {et~al.}(1997){Navarro}, {Frenk}, \&
  {White}}]{1997ApJ...490..493N}
---. 1997, \apj, 490, 493

\bibitem[{{Noble} {et~al.}(2007){Noble}, {Leung}, {Gammie}, \&
  {Book}}]{2007CQGra..24S.259N}
{Noble}, S.~C., {Leung}, P.~K., {Gammie}, C.~F., \& {Book}, L.~G. 2007,
  Classical and Quantum Gravity, 24, S259

\bibitem[{{Papnoi} {et~al.}(2014){Papnoi}, {Atamurotov}, {Ghosh}, \&
  {Ahmedov}}]{2014PhRvD..90b4073P}
{Papnoi}, U., {Atamurotov}, F., {Ghosh}, S.~G., \& {Ahmedov}, B. 2014, \prd,
  90, 024073

\bibitem[{{Perlick} {et~al.}(2015){Perlick}, {Tsupko}, \&
  {Bisnovatyi-Kogan}}]{2015PhRvD..92j4031P}
{Perlick}, V., {Tsupko}, O.~Y., \& {Bisnovatyi-Kogan}, G.~S. 2015, \prd, 92,
  104031

\bibitem[{{Perlick} {et~al.}(2018){Perlick}, {Tsupko}, \&
  {Bisnovatyi-Kogan}}]{2018PhRvD..97j4062P}
---. 2018, \prd, 97, 104062

\bibitem[{{Pratap Singh}(2017)}]{2017arXiv171102898P}
{Pratap Singh}, B. 2017, ArXiv e-prints, arXiv:1711.02898

\bibitem[{{Pratap Singh} \& {Ghosh}(2017)}]{2017arXiv170707125P}
{Pratap Singh}, B., \& {Ghosh}, S.~G. 2017, ArXiv e-prints, arXiv:1707.07125

\bibitem[{{Psaltis} {et~al.}(2015){Psaltis}, {{\"O}zel}, {Chan}, \&
  {Marrone}}]{2015ApJ...814..115P}
{Psaltis}, D., {{\"O}zel}, F., {Chan}, C.-K., \& {Marrone}, D.~P. 2015, \apj,
  814, 115

\bibitem[{{Rahaman} {et~al.}(2010){Rahaman}, {Nandi}, {Bhadra}, {Kalam}, \&
  {Chakraborty}}]{2010PhLB..694...10R}
{Rahaman}, F., {Nandi}, K.~K., {Bhadra}, A., {Kalam}, M., \& {Chakraborty}, K.
  2010, Physics Letters B, 694, 10

\bibitem[{{Saha} {et~al.}(2018){Saha}, {Modumudi}, \&
  {Gangopadhyay}}]{2018GReGr..50..103S}
{Saha}, A., {Modumudi}, S.~M., \& {Gangopadhyay}, S. 2018, General Relativity
  and Gravitation, 50, 103

\bibitem[{{Schee} \& {Stuchl{\'{\i}}k}(2009)}]{2009IJMPD..18..983S}
{Schee}, J., \& {Stuchl{\'{\i}}k}, Z. 2009, International Journal of Modern
  Physics D, 18, 983

\bibitem[{{Spergel} \& {Steinhardt}(2000)}]{2000PhRvL..84.3760S}
{Spergel}, D.~N., \& {Steinhardt}, P.~J. 2000, Physical Review Letters, 84,
  3760

\bibitem[{{Synge}(1966)}]{1966MNRAS.131..463S}
{Synge}, J.~L. 1966, \mnras, 131, 463

\bibitem[{{Takahashi}(2005)}]{2005PASJ...57..273T}
{Takahashi}, R. 2005, \pasj, 57, 273

\bibitem[{{Tsukamoto}(2018)}]{2018PhRvD..97f4021T}
{Tsukamoto}, N. 2018, \prd, 97, 064021

\bibitem[{{Tulin} \& {Yu}(2018)}]{2018PhR...730....1T}
{Tulin}, S., \& {Yu}, H.-B. 2018, \physrep, 730, 1

\bibitem[{{Ure{\~n}a-L{\'o}pez} {et~al.}(2002){Ure{\~n}a-L{\'o}pez}, {Matos},
  \& {Becerril}}]{2002CQGra..19.6259U}
{Ure{\~n}a-L{\'o}pez}, L.~A., {Matos}, T., \& {Becerril}, R. 2002, Classical
  and Quantum Gravity, 19, 6259

\bibitem[{{Vetsov} {et~al.}(2018){Vetsov}, {Gyulchev}, \&
  {Yazadjiev}}]{2018arXiv180104592V}
{Vetsov}, T., {Gyulchev}, G., \& {Yazadjiev}, S. 2018, ArXiv e-prints,
  arXiv:1801.04592

\bibitem[{{Wang} {et~al.}(2017){Wang}, {Chen}, \& {Jing}}]{2017JCAP...10..051W}
{Wang}, M., {Chen}, S., \& {Jing}, J. 2017, \jcap, 10, 051

\bibitem[{{Wei} \& {Liu}(2013)}]{2013JCAP...11..063W}
{Wei}, S.-W., \& {Liu}, Y.-X. 2013, \jcap, 11, 063

\bibitem[{{Xu} {et~al.}(2018{\natexlab{a}}){Xu}, {Hou}, \&
  {Wang}}]{2018CQGra..35k5003X}
{Xu}, Z., {Hou}, X., \& {Wang}, J. 2018{\natexlab{a}}, Classical and Quantum
  Gravity, 35, 115003

\bibitem[{{Xu} {et~al.}(2018{\natexlab{b}}){Xu}, {Hou}, \&
  {Wang}}]{2018arXiv180609415X}
---. 2018{\natexlab{b}}, ArXiv e-prints, arXiv:1806.09415

\bibitem[{{Younsi} {et~al.}(2016){Younsi}, {Zhidenko}, {Rezzolla}, {Konoplya},
  \& {Mizuno}}]{2016PhRvD..94h4025Y}
{Younsi}, Z., {Zhidenko}, A., {Rezzolla}, L., {Konoplya}, R., \& {Mizuno}, Y.
  2016, \prd, 94, 084025

\bibitem[{{Yumoto} {et~al.}(2012){Yumoto}, {Nitta}, {Chiba}, \&
  {Sugiyama}}]{2012PhRvD..86j3001Y}
{Yumoto}, A., {Nitta}, D., {Chiba}, T., \& {Sugiyama}, N. 2012, \prd, 86,
  103001

\end{thebibliography}

\end{document}